\documentclass{elsart1p}

\usepackage{natbib}

\usepackage{graphicx}

\usepackage{amssymb}

\begin{document}

\begin{frontmatter}



\title{Preparing an ESO proposal}



\author[label1]{Pierre Kervella}
\ead{pierre.kervella@obspm.fr}
\author[label2,label3]{Paulo J. V. Garcia}

\address[label1]{LESIA, UMR 8109, Observatoire de Paris-Meudon, 5, place Jules Janssen, 
92195 Meudon, France}
\address[label2]{Centro de Astrof'sica da Universidade do Porto, Rua das Estrelas,
4150-762 Porto, Portugal}

\thanks[label3]{The summer school lecture at Goutelas was given by P.G.}

\begin{abstract}
Preparing observing proposals is a classical exercise in an observational
astronomer's life. In this paper, we summarize the general requirements to write a good proposal
for the ESO telescopes, and eventually obtain telescope time.
We detail in particular the different proposal types, how they are evaluated by ESO,
what they should contain and what to avoid.
The advice given in the present paper are applicable to observing proposals for the
AMBER and MIDI instruments of the VLTI, but also to other ESO instruments,
and to some extent to other observatories.
\end{abstract}


\end{frontmatter}


\section{Introduction}

The first and most important advice of this paper is to {\bf read carefully
the ESO Call for Proposals (hereafter CfP) and the User's Manual for Phase 1 proposals},
that are available at \texttt{http://www.eso.org/observing/proposals}.
Though this may seem obvious, the availability of
a large number of instruments and telescopes results in a complicated set of
rules and limitations that are not necessarily mentioned in the present paper.
The informations are up to date as of period 79 of ESO. Be aware that the rules
for proposal writing and evaluation evolve rather quickly, especially as new instruments
come on line. So keep this in mind and check the ESO web site carefully.
There are two proposal submission deadlines per year at ESO, around 1$^{\rm st}$~April and
1$^{\rm st}$~October (except for Director's Discretionary Time). They correspond to
observations scheduled six to twelve months later.
In this paper, we will describe the different types of proposals
(Sect.~\ref{proposal_types}), how they are evaluated by ESO (Sect.~\ref{opc}), and how to fill he
ESO form (Sect.~\ref{esoform}). A summary of the most important points to remember
is finally given in Sect.~\ref{summary}.

\section{Different kinds of proposals}
\label{proposal_types}

All ESO proposals have in common that they request telescope time in order to
fulfill a scientific goal. However, they come in different types depending on the
amount of time requested, the type of target, or the urgency of the observation.
The CfP defines each programme type in details, and as this definition may
change with time, you are invited to check the corresponding part of this document.

\subsection{Normal and Large programmes}
Normal programmes (NPs) can be defined by default as all the programmes that do
not belong to the other categories decribed below. The time request covers no more
than a single semester (also called ``period"). NPs represent a dominant fraction of all
proposals, as there is little constraints on them. The majority of the available
telescope time is dedicated to the execution of NPs.

%

When a programme requires more than 100\,h of telescope time per semester for
a maximum of four semesters (it can be requested in a single semester), then it qualifies
as a large programme (LP). This type of proposal is typically used for large surveys that
ask for the observations of hundreds or thousands of targets, but it also includes
repeated observations of a small number of objects, or even of a single object. The LPs
can be scheduled up to 30\% of the available telescope time.
While LPs are less often approved by ESO than the NPs, they are interesting
as they benefit from a high scheduling priority at the telescope, and they can be
continued over several periods without the need to ask again for time. As the LPs
requested time is significantly larger than for NPs, the scientific justification has to
be written particularly carefully. As a remark, the applicant to a LP may specify in the 
ESO form that his proposal should be evaluated as a NP if it is rejected by the panel
as a LP. In this case, only the part of the proposal corresponding to the current
period is evaluated.

\subsection{Director's Discretionary Time (DDT) and Target of Opportunity
(ToO) proposals \label{ddt_prop}}

When a sudden and unexpected astronomical event requires an immediate observation,
the Director's Discretionary Time proposals are the best choice. They can also be
obtained when the scientific topic is the subject of a strong competition, for follow-up
observations of another campaign, or to test the
feasibility of a programme. Up to 5\% of all ESO time is available for DDT proposals.
Two important particularities of DDT proposals is that they are evaluated by
a distinct OPC panel (DDTC) from the NPs, and they can be submitted at any time
(there are no DDT deadlines). Another interesting aspect is that approved
DDT proposals have a high priority for execution at the telescope.
Moreover, the acceptance rate of the DDT proposals is higher than
the NPs, with about 50\%. From these advantages, it appears
that {\it if} a scientific programme fits in the relatively restricted framework
of DDT proposals, then it {\it should} in general be proposed as DDT.
As a remark, a DDT proposal already rejected by the OPC panels will, as a
general rule, be rejected by the DDTC.


There is a subtle distinction between the target of opportunity proposals
and the Director's discretionary time (DDT) proposals. ToOs are applicable to
sudden astronomical phenomena that require an immediate observation, but
that are predictable in a generic sense. For instance, supernovae are the typical
case for a ToO proposal, because supernovae are {\it sudden}, but they are not
{\it unexpected}. We can't say when a supernovae will explode, but we know that
a number of them will likely explode in a given period of time.
and therefore they must be prepared in advance, before the expected event.
A specialized {\it rapid response mode} (RRM) is available on some instruments
of ESO (UVES, ISAAC, FORS,...) for very fast acquisition of ToO targets
(typically 6 minutes after the request is received). RRM is particularly useful for
the study of gamma ray burst optical and infrared afterglows. The evaluation
and time allocation of ToOs is done by the same OPC as NPs and LPs,
and they share the same two deadlines per year.

\subsection{Guaranteed time observations, Chilean proposals, OPTICON proposals}
\label{gto}

These three ``subtypes" are in reality not distinct from the NPs, LPs, DDT and ToO,
but they represent different ``political flavours". Guaranteed time observations
(GTO) are generally requested by european teams who were rewarded from the
effort of building an ESO instrument by reserved telescope time. GTO proposals have
a high priority, and though the OPC has in principle the possibility to reject them,
they are in practice always scheduled with a high priority.

The Republic of Chile has, by special agreement with ESO, a right to 10\% of all telescopes.
A ``host state proposal" (also called ``Chilean proposal") is when the Principal
Investigator (PI) is affiliated with a Chilean institution. Chilean proposals are evaluated by
the OPC as others, but they benefit from a special scheduling. Their acceptance rate
is significantly higher than NPs, and a collaboration with astronomers from Chilean
institutes is a good manner to take benefit of this reserved time.

OPTICON is an EU funded network that can finance travel costs to the ESO observatories
for non ESO states nationals (within the EU or associated countries).

\section{The Observing Programmes Committee}
\label{opc}

The Observing Programmes Committee (OPC) is responsible at ESO for the evaluation
and grading of all the proposals (except DDT, see Sect.~\ref{ddt_prop}) that are submitted
each semester (usual deadlines are 1st April and 1st October). The result of the grading
by the OPC is a ranking of all proposals, that eventually governs, through a rather complex
procedure (not discussed in this paper), the way the observations are
scheduled and executed at the observatories.

\subsection{Structure of the OPC}

The OPC is made of four panels who are respectively responsible for proposals related to
cosmology (panel A), galaxies (B), star formation, ISM and planets (C) and stellar evolution (D).
Each of these panels is divided in two or three subpanels each made of five expert advisers
and one OPC member. The subpanels are the groups that actually discuss and grade the
proposals. In total, there are currently 10 subpanels giving a total of 60 astronomers involved
in the evaluation of ESO proposals. The panel members are chosen among the
active astronomers worldwide, excluding the ESO staff. Their participation to an OPC
panel is on a voluntary basis (on request from ESO) and does not give
right to any financial retribution.

\subsection{Proposal evaluation and ranking}

Over the last few periods, ESO has received about 900 proposals per period, and this number
is increasing. As a result, each subpanel receives an average of 90 proposals to evaluate,
slightly more in the C and D panels, and slightly less in A and B. Each proposal is refereed
by three members of the subpanel to whom it was attributed, and read by the three others.
The comments and pre-grading are entered independently by the three referees
in the ESO computer system, and subsequently distributed to all member of the subpanel.
Each observing run of each proposal is graded separately.
All this happens before the OPC meeting, that usually happens six weeks after the
proposal deadlines. The panel members meet in person at this occasion.

During the OPC meeting, each proposal is discussed in the subpanel for an average of
5 to 7 minutes. This can appear as a short time, but in reality the (much longer)
preparation of the meeting serves as a basis for the discussion to focus on the strong
and weak points of the proposal. All six members of the subpanel give a grade to the
proposal at a vote that concludes the discussion. The average of these grades is
used eventually to rank the proposal, or to eliminate it (if really bad). It happens that
conflicts of interest are found at the meeting between the proposal and subpanel
members. In this case, they do not participate to the discussion (they usually leave the
room) and they do not vote.

The large programmes are a specific case and are discussed
by each complete panel, consisting of 12 to 18 persons. The results of the time
allocation are communicated to the PI of each programme via the ESO web letters
system about three months after the proposal submission deadline. Each accepted
proposal is attributed a priority level (A, B or C) based on its ranking.

\subsection{Criteria for a good proposal}

The criteria for the evaluation of the proposals by the OPC panels are clearly defined
by ESO and respected by the panels. They are the following:
\begin{enumerate}
\item The scientific merit of the proposal and its contribution to the
advancement of scientific knowledge.
\item Evidence of sufficient time and resources and of a detailed strategy for a complete
and timely data analysis.
\item The scientific output from previous observations on ESO telescopes (papers published
or in preparation, reports,...).
\item Good prospects of success in terms of achieving the scientific goals described in the
proposal. The feasibility aspects are examined by ESO staff and not by the OPC.
\end{enumerate}
The first criterion is naturally the most important. It is evaluated by the panels
taking into account the requested telescope time. A proposal requesting very little time
with a potentially superb impact will get the highest grade. On the other hand,
a survey that just proposes to add 10 objects to 100 already known, and that asks for
100h of UT time will get a very low rank (or simply be rejected).

\subsection{Caveats}

The best ranked proposals generally obtain telescope time, and the lowest rank
proposals do not. In between these extremes, the situation is not so simple, as
different proposals have different environment and timing constraints
(seeing, transparency,...) that can result in the execution of a low rank
proposal and the non-execution of a better one.

After the OPC meeting, one of the referees (the prime referee) is responsible
for writing the comment card that is sent to the applicant. When reading this card, you
should keep in mind that at the time the card is written, the referee only knows the rank
of the proposal, and cannot be sure that the observations will be executed, or even scheduled.
The proposal evaluation process is complex and can result in apparent
inconsistencies, but one has to keep in mind that the exact amount of ``good seeing"
time is unpredictable.

Each ESO proposal is read by six independent astronomers, and the total
``manpower" devoted to this process, the discussion and the vote is more than two
hours per proposal. Roughly two thirds of the proposals are recommended for
execution, and one third to one half are eventually executed, depending on the pressure
factor. So overall, if a proposal is really good, the chances that it gets rejected are rather low.

\subsection{Pressure factor}

The pressure factor (PF) can be defined as the number of requested nights divided by the
total available time for science. The PF changes significantly with time, depending on the season
(nights are shorter in the austral summer), the right ascension (Galactic plane,
Magellanic Clouds,...), and the availability of new instruments. Orders of magnitude for
the PFs at La Silla and Paranal telescopes are the following:
\begin{itemize}
\item 2 to 4 for the 4m class telescopes at La Silla (more for the 2.2m)
\item 3 to 5 for the 8m Unit Telescopes at Paranal (less for the VLTI ATs)
\end{itemize}
It may be amazing to see that the pressure factor is not very different in practice between
UTs and 4m class telescopes. This can be explained pretty easily however, by the relative
rarity of these smaller telescopes (only 2 telescopes at ESO compared to 4 UTs), that
are well suited for a number of bright objects. So as a general rule, you
should not consider that getting 4m class telescope time is easier than getting UT time.
These two classes of telescopes are suited to different types of objects, but the scientific
value of the proposal should be just as good for both in order to get approved.

\section{How to fill the ESO form}
\label{esoform}

\subsection{Title, category, abstract (boxes 1 and 2)}

The title should be concise to have a maximum impact, and the phrasing should be appealing.
It should summarize the scientific goal and not simply describe the instrumental means that will
be used to reach them, for instance ``NACO observations of active galactic nuclei" is not
informative enough, while ``A high resolution exploration of the torus of three active galactic
nuclei" is better (though it could be more precise). The point is to show clearly the final
astrophysical goal of the study. 

There is a two-characters code to define the OPC panel to which your proposal will go.
Be careful to choose the right category in order to get the most competent experts to
review your proposal. There are some ambiguous objects that can be classified in
two categories (e.g. brown dwarfs/giant exoplanets).
In this case, think about the precise goal of the observations, and you
should be able to choose the right category and sub-category.

The abstract is very important because it should raise the interest
of the reader and show him the context and added value of the proposed observations.
It should clearly delimit the main objectives of the proposal, before going into
further details in pages 2 and 3. As the available space in the frame is short,
take care of the phrasing in order not to overfill the box.
As a remark, you should not
use a smaller font than the default one (this is true for the whole ESO form)
to gain space artificially, and you should not use any LaTeX symbol in the title
and abstract (use ``alpha" instead of ``$\alpha$" for instance). This is written about
a hundred times in the ESO form, but a significant fraction of the proposals still
contain such typos. To check if the title and abstract are good, ask yourself
if they answer to the questions ``Why (context, scientific problem), How (instruments/object),
What (final data/knowledge)".

\subsection{Observing runs (box 3)}

In the ESO terminology, a proposal is requesting observing runs. They are defined as
a block of instrument time with given constraints in terms of observing conditions
(seeing, transparency), time of the year, service or visitor mode,... A run can only concern
a single instrument and setup, but you can split a request on one instrument
into several runs. In other words, you can ask for several runs with the same instrument/setup,
but not for several instruments grouped in a single run.

So, why ask for several runs instead of one for a single instrument setup ?
For VLTI observations, one run can only be defined with one baseline,
so a combination of different baselines requires the definition of multiple runs.
In visitor mode, the separation in observing runs is rather clear. One trip to Chile usually
corresponds to one observing run. 
In service mode, the separation can be arbitrary, though it is traditionally one run per
instrument. If you ask for 20 stars to be observed with AMBER for instance,
you can ask for one single run (standard way), split these observations in two runs of 10 stars,
or even, in theory, 20 runs on one star each (there is a maximum of 26 runs allowed in the
ESO form). The only advantage to spllit runs is that this gives more flexibility to the OPC panel
to grade each run separately. The typical case to split an observation in two runs is if you
have first and second priority stars, grouped in two separate runs.

If the OPC is not convinced that your science case is strong enough to allocate
time for all stars, he still has the possibility to cut only run B. The OPC cannot reduce
the time allocated to one run compared to what is requested, except in exceptional
cases. This means that a run is always ranked as a whole. Be careful however not to
split too much into ``subruns": you should not rely on the OPC to select your targets...

This is also in this box that you specify the environmental constraints of your runs (maximum
seeing, transparency, season). You should set these numbers to the least constraining
values that are acceptable, in order to maximize your chances to be scheduled.
The median seeing at Paranal is of the order of 0.8\,arcsec, thus if you ask for this seeing, your
observations will be executable only half of the time.

\subsection{Already allocated and future time, special remarks (boxes 4 and 5)}

The usual way of getting time over several ESO periods is to submit a large programme.
However, in some cases, the requested time is too short to qualify for this type of proposal,
or the time request for a given period depends on the result of the preceding period.
In this case, the only way is to ask for time several times using normal proposals.
As a general rule, you should not forget to mention the time allocated by ESO in the
past to your programme. This is usually checked by the referees, and if you do not
mention it, this will result in a downgrading of your proposal. There may be very good
reasons why you ask again for time, just explain them ! There is a short space in box 5
that can be used for this purpose, and you can continue in pages 2 and 3 if necessary.
A good reason to fill this frame is for instance if you experienced a partial or total loss
of your approved time due to bad weather.

You can be less exhaustive regarding future observations still required to
complete the project, especially if they depend on the results of your current proposal.
However, if for instance you need objects that are not observable in the current
period for a survey, do fill the corresponding line, as it will be received positively
by the OPC as the indication of a careful planning.

The special remarks box can be used in practice for anything (special instructions,
scheduling constraints,...) that should be known by the OPC. For ToO and RRM
proposals, the number of triggers should be mentioned here.

\subsection{Investigators and PhD students (boxes 6 and 7)}

The choice of the PI and co-Is should be agreed clearly within the team. One should
remember that the PI does not need to be the person who writes the paper
that will come out of the observations. The nationality and affiliation
of the PI is not an issue with respect to the proposal selection
process (except for Chilean time proposals, see Sect.~\ref{gto}).

There is a special box to indicate the presence of a PhD student in the team.
This is considered by the OPC as an indication that
the data will be analyzed and published quickly. Students are usually pressed by deadlines,
such a a thesis defense, and this helps getting papers published in time. So, if there is a
student in the project, do not forget to mention it, as this will raise the proposal
in the ranking (though not to the point of making a bad proposal get time).
First time proposals from PhD students are particularly encouraged.

\subsection{Scientific rationale (box 8A)}

This is naturally the most important part of the proposal. This is where you should
demonstrate that the telescope time you are asking for will actually lead to a progress
of scientific knowledge. The scientific rationale should be written in a similar form
to a paper, however with a simpler introduction as it will be read by a panel of
non-specialists. A good description of the state of the field at large is very welcome.

Your text should show what is currently missing in the global picture of astrophysics
and that you will uncover with the requested observations. For instance, if you
want to detect and measure the angular size of a circumstellar disk using AMBER, you
should explain why such a disk is expected (other observations, literature,...) and
what this measurement will tell us about its physics. You should also discuss the
potential problems and limitations of the observations to adress the scientific
question (e.g. will the limb darkening or flattening of the star
be a problem to detect the disk?).

You should use short sentences and give a structure to your text.
A block of text without sections is particularly difficult to read, and
will have a negative impact on the apreciation by the OPC.
You should use bold face titles, section numbering,... just as you would
in a journal paper.

References are important to position your research in the field, but they should
be kept to a limited number for two reasons.
Firstly, it is unlikely that the referees will actually read them, or even know them
(remember that the OPC panel will usually {\it not} have a specialist of your field).
Your text should be clear enough to be understandable by itself.
Secondly, they take space in the box,
and this is at the expense of explanations of your programme. A compact
way of listing the references is to group them at the end of the box 8A on a single
line, with a single author name in bold font (use ``et al." for multiple authors),
and separated by ``;". You can also write the references directly in the text,
as it may be easier to read.

Figures are important and valuable, so add as many useful figures as
possible within the two pages allocated for this purpose.
For instance, if you want to observe an emission line with AMBER, insert a
spectrum to actually show this line. A picture of the model you plan to use is also
a good way of showing the level of detail you are aiming at.
You should quote the figures in the text using
bold characters, as this facilitates the back and forth reading of the figures and
text (they are located on different pages in the form).

\subsection{Immediate objective (box 8B)}

This section should be used to explain what the observations will actually measure
physically, and how the results of these measurements will be converted into
useful scientific knowledge. For instance, considering the case of a circumstellar
disk observation with AMBER, this box should tell how the visibilities
and closure phase will be translated into the angular size, the
flux contribution and the asymmetry of the disk, and how this will be used to
estimate its mass and temperature profile.
For interferometric observations, this section is
a good place to describe the model that will be fitted to the data.

\subsection{Telescope and observing mode justification (boxes 8C and 8D)}

For VLTI observations, the telescope justification is straightforward, as this is the
only interferometric facility open to European astronomers. For other telescopes
and instruments, be careful however to explain why your programme needs
ESO telescopes and cannot be done at other observatories. Some instruments
are available on both the UTs and the 4m class telescopes of ESO, such as
ISAAC (UT)/SOFI (NTT), or FORS (UT)/SUSI (NTT). You should justify why you
need the big telescopes.

The observing mode (service or visitor) can appear as a detail, but it is not.
For ESO, the scheduling is very different for the two types of observations,
and for you it means different levels of weather risk. In short, visitor mode (VM) should be
requested for any observation that requires an interaction at the telescope.
It is also required for observations in some special instrument modes.
All ESO telescopes are available in VM, and an advantage of VM is that
you can really optimize the setup to get the highest possible data quality.
Additionally, you get a plane ticket to Chile for free (if you are from an ESO
member country). The risk of VM for you is that you may get
bad weather during part of your run, or even completely lose it.
The dates are fixed a long time in advance and cannot be changed,
so if clouds cover Paranal during your night, you will not get any data
nor any compensation time for this loss. As a remark, it can happen
that ESO changes the observing mode from VM to SM or vice versa during
the scheduling process.

Service mode (SM) is much more flexible. If you choose SM, the observations
will be executed for you when the atmospheric conditions
are within the limits you have specified in your proposal. If the conditions
degrade during the observation, it will be repeated (provided it got a sufficiently
good rank from the OPC). The data are sent to you a few weeks after the observation.
The big advantage of SM is that the risk for bad weather is partially covered by ESO.
However, if your constraints are too strict, your programme may still not be executed.
The disadvantage of SM is that you will obtain exactly the data that you requested
in your proposal, nothing more. If you selected the wrong instrumental setup,
you will only see it when you receive the data, and no compensation will be given
by ESO. But if your observations are straightforward to prepare and easy to execute,
SM is the best choice.

There are minimum times specified in the ESO Call for Proposal of 2 nights
for VM and 6 hours for SM. These limits should not be considered ``hard",
and provided your proposal is good, you will get time even if you requests less
time than that.

\subsection{Strategy for data reduction and analysis (box 8E)}

You should explain there how you will practically reduce and interpret the data, what
are the domains of expertise of the members of the team, and state the available
manpower to process the data. You do not need to go into too much details, but
your overall method should appear clearly. For AMBER and MIDI, you can quote
for instance the data reduction software that you will use. If you use several
techniques (interferometry, spectrography, photometry,...), you should give
details on the processing of each data set, and how they will be combined together.

\subsection{Time justification (box 9)}

The accounting of the minimum necessary hours should be done precisely
using the ESO exposure time calculators (ETCs) that are available on the
web, and the instrument manual to estimate the overheads. You should not add
artificially some time to correct for bad weather. You should not round up
the time you obtain too loosely. When specifing the target SNR
in the ETC, do not ask for too much, but really what you need to fulfill the science
goal of your proposal. If you ask for too much time, the OPC will downgrade
your proposal. If you ask for too little time, it will also be downgraded as the
feasibility of the programme will be doubtful. Telescope time estimation
is a delicate exercise.

The lunar phase is in general irrelevant for infrared observations, especially
interferometric observations that can be done in bright time.
If you need specific additional calibrations, you should specify
it by activating the ``special calibration" keyword in the ESO form, and add the
corresponding time to your request.

\subsection{Report on previous use of ESO telescopes (box 10)}

This is a special box to describe the previous use of ESO telescopes,
and it is important to fill it carefully. If you have received a lot of time in the past and no papers
have been published out of them, this is a negative point (though not disqualifying).
If you are an experienced ESO user with a good record of publications, the
probability that you get time will be higher.

\subsection{Applicants' publications related to the proposal (box 11)}

One of the criteria for the evaluation of proposals is the prospects for a timely
and complete analysis of the acquired data. This is checked by the OPC using among other
things the publication record of the members of the proposing team. Many publications on a subject
are in general a positive point, but teams who are starting a research project on a new subject
are not penalized.

\subsection{List of targets and ESO archive (boxes 12a and 12b)}

The list of targets is straightforward to establish, but do not discuss scientific details
in the target notes part. As a general rule, the referees {\it will} check
the availability of the data corresponding to your objects in the
ESO archive (box 12b). If they find that the observations you request have already been
done, and you do not mention it here, this will be a case for rejecting your proposal,
or grading it very low. So take a few minutes to make a search in the ESO archive,
as you may find the data you need without even submitting the proposal!
Even if some data is in the archive, you can very well argue that it is not sufficient
or not suitable for your project, but you have to discuss it here.

\subsection{Scheduling requirements and instrument configuration (boxes 13 and 14)}

The scheduling box is rarely used presently for interferometry, as the availability
period of each baseline is not known at the time of the submission of the proposal.
If you ask for visitor mode observations, you can specify the dates
that are unsuitable for you to go to Chile.
If you have simultaneous observations with other telescopes, you can indicate
the dates here. The repetition of observations of changing sources is also
described here.

The instrument configuration (box 14) is described using codes. You must read carefully
the manual of the instrument in order to select the optimal setup for your
programme. As a remark, the overheads often depend on the setup chosen, though
for the moment, AMBER and MIDI observations have a fixed observing time per target.
ESO instruments are very versatile and, for most of them, have a vast number
of filters, spectral resolutions,... You should be really familiar with your instrument
to get the most from it.

\subsection{List of interferometry targets (box 15)}

This box can seem redundant with box 12a, but in reality it adds useful information
for the scheduling at ESO, and the assessment of the feasibility.
In particular, this is here that you specify the baseline you want to use, and the
expected visibility $V$ of the fringes. The correlated magnitude
$m_C$ of the target is defined as $m_C = m - 2.5 \log_{10}(V)$, and you can use the ASPRO tool
of the JMMC to compute it. As a quick check, it should always be larger than the regular
magnitude. There are limiting magnitudes for AMBER and MIDI in the CfP, that are expressed
in correlated magnitudes. If your object has a higher magnitude than these limits, the observation
may be judged unfeasible, unless you demonstrate otherwise.

\section{Some things to remember}
\label{summary}

\begin{itemize}

\item {\bf Be very clear in explaining your goals.}
Limit acronyms to the minimum and use understandable words.
The phrasing should be absolutely crystal clear for you,
if you expect it to be readable by non-specialists. The referees are
usually not in your field (though they probably know vaguely something
about your subject). A referee who does not understand a proposal will
give at best an average grade. Due to the high pressure factor, it is likely that
it will not be executed. Try to have other people read it,
especially who are not working in the same field as you do (consider for instance a
planetologist if you study black holes). {\it Do not take for granted that there will be
an expert of your field in the OPC panel.} For VLTI proposals in particular,
you should clearly explain why and how interferometry can add scientific information,
without boring the referee with technical details.

\item {\bf Evaluate the ratio scientific return/telescope time.}
Huge programmes with a low scientific return make a bad use of expensive telescope
time. If you need a number of targets to establish statistical properties, this can be
a good reason, but beware of the ``doubling the sample" syndrom: if 100 targets
of a kind are already known, addding another 100 through tedious observations over
hundreds of hours will only result in reducing the uncertainty by (at best) $\sqrt{2}$
on a model parameter. In other words, statistics are usually expensive in
telescope time, and this is even more the case for interferometric observations.
Even if you focus on a reduced number of targets, be very clear in your estimate of
the telescope time you need. Do not round up the time too loosely: one night of VLTI+UTs
corresponds to tens of thousand euros...

\item {\bf Do not hesitate to propose innovative and ambitious proposals.}
The OPC likes very much the proposals that can have a high impact
on their field, even if they may give no result at all. So if your object is very difficult
to observe,  but the science is very exciting, go ahead and propose the observation,
there is a good chance that it gets approved. You can also think about submitting a
DDT proposal to demonstrate the feasibility of the project.

\item {\bf Do not be too specific and avoid irrelevant details.}
A proposal is a different exercise from a refereed journal paper. You do not have to
prove in details every bit of your reasoning. Your goal is to obtain time, and you should 
not divert too much from this objective. Your proposal should convey the feeling that
you have a clear and broad view of your subject, and that you will know what to
do with the data you will obtain. Going into technical details can confuse the referees
and will certainly not give them the impression that you are an expert.

\item {\bf Do not ask for too stringent conditions.}
Except if your object is at the limit of the possibilities of the instrument, you should
not ask for extraordinary seeing and transparency conditions, or your programme
may not be scheduled.

\item {\bf Read at least twice your proposal after you finished writing it.}
This will allow you to remove all the typos, correct the phrasing, and improve the text layout.
A proposal full of typos has usually a negative {\it a priori} against it. Remember that many
people will spend time reading it, and if they feel that they spent more time doing so
that it took you to write it, this is a problem.

\item {\bf Give a structure to your scientific rationale.}
Use LaTeX properly, including bold and italic fonts, but {\it never} reduce the font size.
This would give the rather bad impression that you are cheating compared to other
proposals.

\item {\bf Check the figures.}
They should be easily readable in a black and white laser print. Not all referees
have color printers in their institutes, and if the figures are unreadable, this will
weaken your argumentation.

\item {\bf Do not submit too many proposals.}
The number of proposals you submit should not exceed your capacity to process
and publish the data. Being overwhelmed with data is definetely not a
comfortable situation.

\item {\bf Do not get discouraged if your proposal is not approved.}
Use the comments from the web letters to improve your proposal, but do not take them
personally. Remember that they are written by the prime referee of your proposal
at a time when he/she does not know if it will be scheduled or not. So unfortunately,
a positive comment can still correspond to a proposal that will not get time, due to
oversubscription or just to bad weather/tight constraints. The OPC messages
are deliberately short, neutral and general to avoid polemic and useless critique.
The quartile of your proposal (its aproximate rank in the list of all proposals) will
be given starting in P79. When you have the OPC comments, read your proposal
again, and ask your colleagues to give you their feedback. You can also contact
the OPC representative of your country, or ESO through the visiting astronomers
department. And always remember that there is no conspiracy against you!

\item {\bf If you get approved in category A or B but you get no data.}
If ranked in category A (the best) and in service mode, your programme will most
of the time be carried over to the next period and your objects will be observed without
the need for another proposal (except if it was time-critical to occur on
a certain date, for instance). In visitor mode, even if your proposal was in category
A and you receive no data due to weather or technical downtime, you will generally
have to apply again.
In category B, your programme will normally not be carried over, but you
can submit it again with a special remark in the first page of the proposal.
It will have a favorable impact on the appreciation of the OPC. You can also
relax the observing constraints at this occasion.

\end{itemize}

\section{Conclusion}

Writing a proposal is a formal exercise that is both similar and different from
writing a journal paper. It is similar in the sense that you have to expose clearly
your ideas to convince a panel of referees that your programme is scientifically
valuable. But it is fundamentally different because this programme is naturally
not yet done! So you have to be positive and ambitious in your text, though
still objective about the chances of success. Another diffference is that the
referees are generally not experts in your narrow field of research.
But all these efforts are worth it, as successful observations are the key that
led to almost all the most important discoveries in astrophysics.

\section{Further reading}

You will find additional information in the following
web pages and publications:

\begin{itemize}
\item ``Proposal Preparation and Submission" page at ESO:

\texttt{http://www.eso.org/observing/proposals/}

\item Instrumentation at ESO:

\texttt{http://www.eso.org/instruments/}

\item Waelkens, Ch., ``On the writing of observing proposals" (2005)

\texttt{http://www.eso.org/observing/proposals/writing-op.html}.

\item Walsh, J., ``On HST proposal writing", ST-ECF Newsletter, 34, 5 (2003)

\texttt{http://www.spacetelescope.org/about/further\_information/}

\texttt{\ \ \ \ newsletters/pdf/newsletter\_34.pdf}
\end{itemize}









\end{document}